\documentclass{article}
\usepackage[T1]{fontenc} 
\usepackage[utf8]{inputenc} 
\usepackage{ismir,amsmath,cite,url}
\usepackage{graphicx}
\usepackage{color}
\usepackage{amssymb}
\usepackage{amsfonts}
\usepackage{graphicx}
\usepackage{makecell}
\usepackage{adjustbox}

\title{MMT-BERT: Chord-aware Symbolic Music Generation Based on Multitrack Music Transformer and MusicBERT}

\multauthor
{Jinlong Zhu$^1$ \hspace{1cm} Keigo Sakurai$^1$ \hspace{1cm} Ren Togo$^2$} { \bfseries{Takahiro Ogawa$^2$ \hspace{1cm} Miki Haseyama$^2$}\\
  $^1$ Graduate School of Information Science and Technology, Hokkaido University, Japan\\
$^2$ Faculty of Information Science and Technology, Hokkaido University, Japan\\
{\tt\small mhaseyama@lmd.ist.hokudai.ac.jp, jinlong@lmd.ist.hokudai.ac.jp}
}

\def\authorname{Jinlong Zhu, Keigo Sakurai, Ren Togo, Takahiro Ogawa and Miki Haseyama}

\begin{document}

\maketitle
\begin{abstract}
We propose a novel symbolic music representation and Generative Adversarial Network (GAN) framework specially designed for symbolic multitrack music generation.
The main theme of symbolic music generation primarily encompasses the preprocessing of music data and the implementation of a deep learning framework. 
Current techniques dedicated to symbolic music generation generally encounter two significant challenges:
training data's lack of information about chords and scales and the requirement of specially designed model architecture adapted to the unique format of symbolic music representation.
In this paper, we solve the above problems by introducing new symbolic music representation with MusicLang chord analysis model.
We propose our MMT-BERT architecture adapting to the representation. To build a robust multitrack music generator, we fine-tune a pre-trained MusicBERT model to serve as the discriminator, and incorporate relativistic standard loss.
This approach, supported by the in-depth understanding of symbolic music encoded within MusicBERT, fortifies the consonance and humanity of music generated by our method. 
Experimental results demonstrate the effectiveness of our approach which strictly follows the state-of-the-art methods.
\end{abstract}
\section{Introduction}\label{sec:introduction}
Music plays an indispensable role in our daily lives, and there is a significant demand for creating new musical contents. 
Automatic music generation is one of the most intriguing tasks in bringing new music experiences to consumers~\cite{liu2023literature}. 
The earliest studies in the 1950s focused on a combination of music theory and Markov-chains-based probabilistic models, and realized randomly creating music parts and combining them into a synthesis~\cite{Brooks1957AnEI}. 
Contemporary studies have achieved higher quality and faster music generation by utilizing advanced neural networks such as Generative Adversarial Networks (GANs), Transformer, and diffusion models~\cite{goodfellow2014generative, vaswani2017attention, ho2020denoising}. 
Despite significant advancements, previous methods continue to suffer from challenges such as insufficient data extraction and unstable training trajectories. 
Consequently, there is room for new approaches for more effective music representations and more robust deep learning architectures.

In particular, chords are crucial for conveying emotional and humanistic expressions in music, yet few methods take chords into account in symbolic music representation.
Consequently, previous methods are deprived of indispensable information about chords and scales. This lack results in the generation of music that exhibits a diminished degree of humanity.
Therefore, previous methods for music generation face limitations in their ability to produce human-like and high quality expressions~\cite{dong2023multitrack}.
A feasible solution to overcome this difficulty is the integration of a chord analysis model~\cite{huang2020pop}. 
Chord analysis model aids in the extraction of chord data from raw audio, fostering a novel representation method that encompasses chord information~\cite{takuya1999realtime,park2019bi,scholz2008cochonut,demirel2019automatic}. 
With the aid of state-of-the-art chord analysis models, we can generate more harmonious and structured music with more regular chord progressions by automatically extracting and encoding chords from raw audio files.
Therefore, it is expected that adopting chord analysis models in creating new symbolic representations of music will enable the generation of music that is closer to human composition.

Another problem arises from the ever-changing format of symbolic music representation, which makes designing the model's architecture that fits symbolic music generation to be another challenge.
GANs are widely applied in the symbolic music generation field because the addition of a discriminator obviously strengthens the fidelity of the overall generative model~\cite{jhamtani2019modeling,zhang2021implement,walter2021midipgan,muhamed2021symbolic,dong2018musegan,dong2018convolutional}.
The performance of GANs is deeply influenced by the architecture of the generator and discriminator.
Previous studies have demonstrated the effectiveness of transformer-based generators~\cite{hsiao2021compound,dong2023multitrack,huang2020pop,copet2024simple,wu2020jazz,ens2020mmm,yu2022museformer,von2022figaro}. 
Whereas, the architecture of the discriminator has been extensively discussed in recent years. 
Some methods~\cite{dong2018convolutional,jhamtani2019modeling,dong2018musegan,walter2021midipgan} involve constructing a discriminator based on CNN or Transformer, while others~\cite{muhamed2021symbolic} utilize pre-trained models adapted to their tasks.
Compared to hand-crafted discriminators, using pre-trained models often achieves a fairly good result because pre-trained models are already trained on large and diverse datasets. Therefore, the application of pre-trained models will allow the GAN to leverage their learning and knowledge, ensuring the training efficiency and stability.

However, there are few choices of pre-trained models designed for symbolic music representation that can be used as the discriminator considering the input format and length limitation. We solve this problem by applying BERT-based scores, which are well correlated with human ranking and can jointly measure quality and diversity~\cite{salazar2019masked,montahaei2019jointly}. 
Since BERT is trained using a self-supervised loss on bidirectional contexts of all attention layers, it can effectively extract representations~\cite{dai2019transformer, zhang2019bertscore, DBLP}. Muhamed \emph{et al.} employ a pre-trained Span-BERT model and achieve considerable results on harmonic choices and overall music quality, showing that pre-trained BERT-based models outperform CNN-based discriminator~\cite{muhamed2021symbolic}. 
Hence, employing a pre-trained model as a discriminator can amplify the overall performance of the GAN model.

In this paper, we propose a novel symbolic music generation method using the chord-aware symbolic music representation and MusicBERT-based discriminator.
In terms of symbolic musical expression, we introduce the novel symbolic music representation with chord information derived from MusicLang\footnote{https://musiclang.github.io/tokenizer/}, one of the state-of-the-art chord analysis models. 
By employing symbolic music representation with chord information, our model can achieve the generation of more human-like music that considers chord progressions.
For the model architecture, we employ the Multitrack Music Transformer (MMT)~\cite{dong2023multitrack} as the generator and fine-tune the MusicBERT~\cite{zeng2021musicbert}, a symbolic music understanding model pre-trained in large-scale dataset, as the discriminator. Leveraging the superior comprehension capabilities of MusicBERT, we can improve GAN's performance, thereby facilitating the creation of higher-quality music.
Furthermore, we introduce relativistic standard loss to further optimize the stability and consistency of the training process~\cite{jolicoeur2018relativistic}. The use of Relativistic Standard GAN (RS-GAN) has realized great results in the field of image generation. It enables models to account for the fact that half of the data in a mini-batch is fake, leading to more accurate estimations of data realism~\cite{walter2021midipgan, li2023algorithm}.
Building upon the innovations mentioned above, our model is capable of retrieving substantial information about chords and scales, acquiring knowledge in music theory, and autonomously generating multitrack music of superior quality and enriched with human-like characteristics.

The contributions of this paper are summarized as follows.

\begin{itemize}
    \item We propose a modified MMT style symbolic music representation including chord and scale information.
    \item We develop MMT-BERT, an optimized GAN architecture utilizing MMT and MusicBERT, with relativistic standard loss to enhance the stability of the training process and achieve better results.
\end{itemize}

\section{Related Works}\label{sec:Related Works}

\subsection{Symbolic Music Representation}\label{subsec:SMR}
To enable computers to properly understand music, research on symbolic music representation has been conducted for many years~\cite{briot2017deep}.
Musical Instrument Digital Interface (MIDI) is the most commonly used format for symbolic music representation, containing performance data and control information for musical notes.
In the music processing community, many researchers symbolize music with MIDI-like events~\cite{ji2020comprehensive}.

Huang \emph{et al.} have proposed REvamped MIDI-derived events (REMI), which adds note duration and bar events, enabling models to generate music with subtle rhythmic repetition~\cite{huang2020pop}.
However, the REMI representation often encounters a challenge that the sequence is too long.
Building upon the REMI framework, Hsiao \emph{et al.} have proposed Compound Word Transformer (CP)~\cite{hsiao2021compound}. 
CP modifies REMI's approach by transforming one-dimensional sequence tokens into compound words sequence using specific rules. 
Although this modification significantly shortens the average token sequence length and simplifies the model's ability to capture musical nuances, CP is hard to generate multitrack music~\cite{dong2023multitrack}.
Dong \emph{et al.} have proposed their multitrack music representation, which represents music with a sequence of sextuple tokens, along with a Transformer-XL-based generation method Multitrack Music Transformer (MMT).
This approach utilizes a decoder-only Transformer architecture, adept at processing multi-dimensional inputs and outputs. MMT leverages the advantages of the Transformer to enable the generation of longer multitrack music compositions than previous music generation methods. 
However, MMT's representation scheme lacks chord event inclusion, an essential element in musical compositions.
In contrast, our symbolic music representation technique builds on the foundation laid by MMT by integrating chord information, enabling our model to produce more harmonically rich compositions.

\begin{table}[t]
  \begin{adjustbox}{scale=0.75}
    \begin{tabular}{l c c c c} \hline
      Representation & Multitrack & \makecell[c]{Instrument\\control} & \makecell[c]{Compound\\tokens} & \makecell[c]{Chord\\awareness}   \cr \hline 
      REMI\cite{huang2020pop} &  &  &  & \checkmark \cr
      MMM\cite{ens2020mmm} & \checkmark &  &  &  \cr
      CP\cite{hsiao2021compound} &  &  & \checkmark & \checkmark \cr
      FIGARO\cite{von2022figaro} & \checkmark &  &  & \checkmark  \cr
      MMT\cite{dong2023multitrack} & \checkmark & \checkmark & \checkmark &  \cr \hline \hline 
      MMT-BERT (ours) & \checkmark & \checkmark & \checkmark & \checkmark \cr \hline
    \end{tabular}
  \end{adjustbox}
  \caption{Comparisons of related representations.}
  \vspace{-0.2cm}
  \label{comp_representation}
\end{table}

\subsection{Generative Adversarial Network-based Music Generation}\label{subsec:GANArchitecture}
Previous studies have employed various GANs to realize symbolic music generation~\cite{dong2018convolutional, jhamtani2019modeling, zhang2021implement}.
In early states, Dong \emph{et al.} have proposed MuseGAN, a CNN-based GAN architecture, managing to generate multitrack music pieces~\cite{dong2018musegan}.
However, CNN-based GANs often suffer from problems such as limited local perception, fixed-size inputs, etc.
Muhamed \emph{et al.} solved this problem by introducing their Transformer-GANs model, using a Transformer-XL-based generator and pre-trained Span-BERT as the discriminator~\cite{muhamed2021symbolic}. 
Transformer-XL introduces the notion of recurrence into the deep self-attention network, enabling the reuse of hidden states from previous segments as memory for the current segment, allowing for the modeling of long-range dependencies. 
For the discriminator, Span-BERT is utilized to extract sequence embeddings followed by a pooling and linear layer. 
The bidirectional transformer has a comparable capacity to the transformer-based generator and uses the self-attention mechanism to capture meaningful aspects of the input music sequence. 
Their research validates the efficacy of employing a Transformer-XL-based generator in conjunction with a BERT-based discriminator~\cite{neves2022generating}.
Building on this concept, we developed the MMT-BERT model by utilizing MMT as the generator and Music-BERT as the discriminator.

\section{Methodology}\label{sec:Method}

\begin{table}[t]
  \begin{adjustbox}{scale=0.87}
    \begin{tabular}{l c c c c} \hline
      Event type $t_{j}$ & Quintuple token $\mathbf{x}^{t_{j}}_{i}$ \\ \hline 
      start-of-song & (\textsf{0}, \textsf{0}, \textsf{0}, \textsf{0}, \textsf{0}) \\
      instrument & (\textsf{0}, \textsf{0}, \textsf{0}, \textsf{0}, \textsf{instrument}) \\
      start-of-score & (\textsf{0}, \textsf{0}, \textsf{0}, \textsf{0}, \textsf{0}) \\
      note & (\textsf{beat}, \textsf{position}, \textsf{pitch}, \textsf{duration}, \textsf{instrument}) \\
      chord & (\textsf{beat}, \textsf{degree}, \textsf{root}, \textsf{mode}, \textsf{extension}) \\ 
      end-of-song & (\textsf{0}, \textsf{0}, \textsf{0}, \textsf{0}, \textsf{0}) \\ \hline
    \end{tabular}
  \end{adjustbox}
  \caption{The elements of the quintuple token $\mathbf{x}^{t_{j}}_{i}$ for each event type $t_{j}$.}
  \vspace{-0.2cm}
  \label{quintuple_token}
\end{table}

\subsection{Proposed Symbolic Music Representation}\label{subsec:SMR-method}
In our approach, we introduce a novel symbolic music representation that incorporates chords.
\tabref{comp_representation} shows differences between the conventional approaches and the proposed symbolic music representation.
While most conventional representations omit details about chords, we focus on chord information-aware representation to facilitate the process of generating music that more closely resembles humans.
First, we extract music data including chords and notes from MIDI files based on MuspyToolkit~\cite{dong2020muspy} and MusicLang. 
During the extraction process, we recognize a chord once per bar, i.e., every four beats. 
We exclude songs with a time signature other than 4/4, limit the number of chords in a bar to one, and ignore chord changes within a bar since MusicLang only detects chord changes once per bar.
Each time MusicLang detects a chord change, it extracts the scale degree, tonality root, tonality mode, chord octave and extension note of the chord.
After extracting chord and note information, we encode a piece of music into a sequence of quintuple tokens ${\mathbf{X} =(\mathbf{x}_{0},...,\mathbf{x}_{N-1})}$, where ${\mathbf x_{i}}$ and ${\it N}$ denote the ${i}$-th quintuple token and the total number of quintuple tokens, respectively.
Here, ${\it t}$ represents the following event type: \{start-of-song, instrument, start-of-score, note, chord, end-of-song\}.
The meanings of each event type are shown as follows:
\begin{itemize}
    \item \textbf{Start-of-song}: The beginning of the music piece
    \item \textbf{Instrument}: An instrument used in the music piece
    \item \textbf{Start-of-score}: The beginning of a sequence of musical events, including notes and chords
    \item \textbf{Note}: A note characterized by beat, position, pitch, duration, and instrument 
    
    \item \textbf{Chord}: A chord characterized by beat, scale degree, root note, mode, and extension note 
    \item \textbf{End-of-song}: The end of the music piece
\end{itemize}

\begin{figure}
 \centerline{
 \includegraphics[width=1.00\columnwidth]{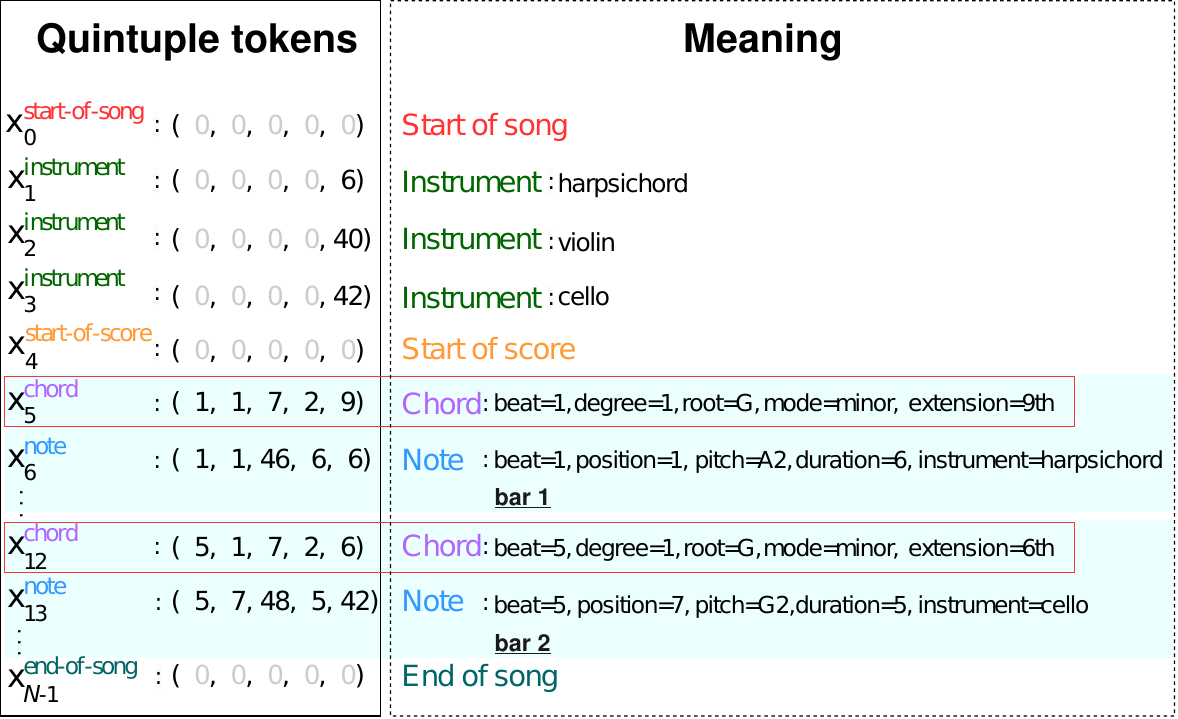}}
 \caption{An example of the proposed representation. Compared to the conventional representation, the proposed representation incorporates an additional chord event (highlighted by red blocks) per bar, thereby aiding the model in understanding the relationship between the notes and chords.}
 \label{fig:repre}
 \vspace{-0.1cm}
\end{figure}

The meaning of each element in the quintuple token $\mathbf{x}^{t}_{i}$ varies depending on the event type ${\it t}$. 
The correspondence between the event type ${\it t}$ and the meanings of the quintuple token $\mathbf{x}^{t}_{i}$ is shown in~\tabref{quintuple_token}.
Additionally, it is noted that we apply different embeddings for the different features sharing the same axis.
A schematic diagram of the proposed representation is illustrated in~\figref{fig:repre}.
In this way, we can obtain a symbolic music representation that incorporates chords that is suitable for input into the aforementioned MMT-BERT architecture.

\begin{figure*}
 \centerline{
 \includegraphics[width=1.5\columnwidth]{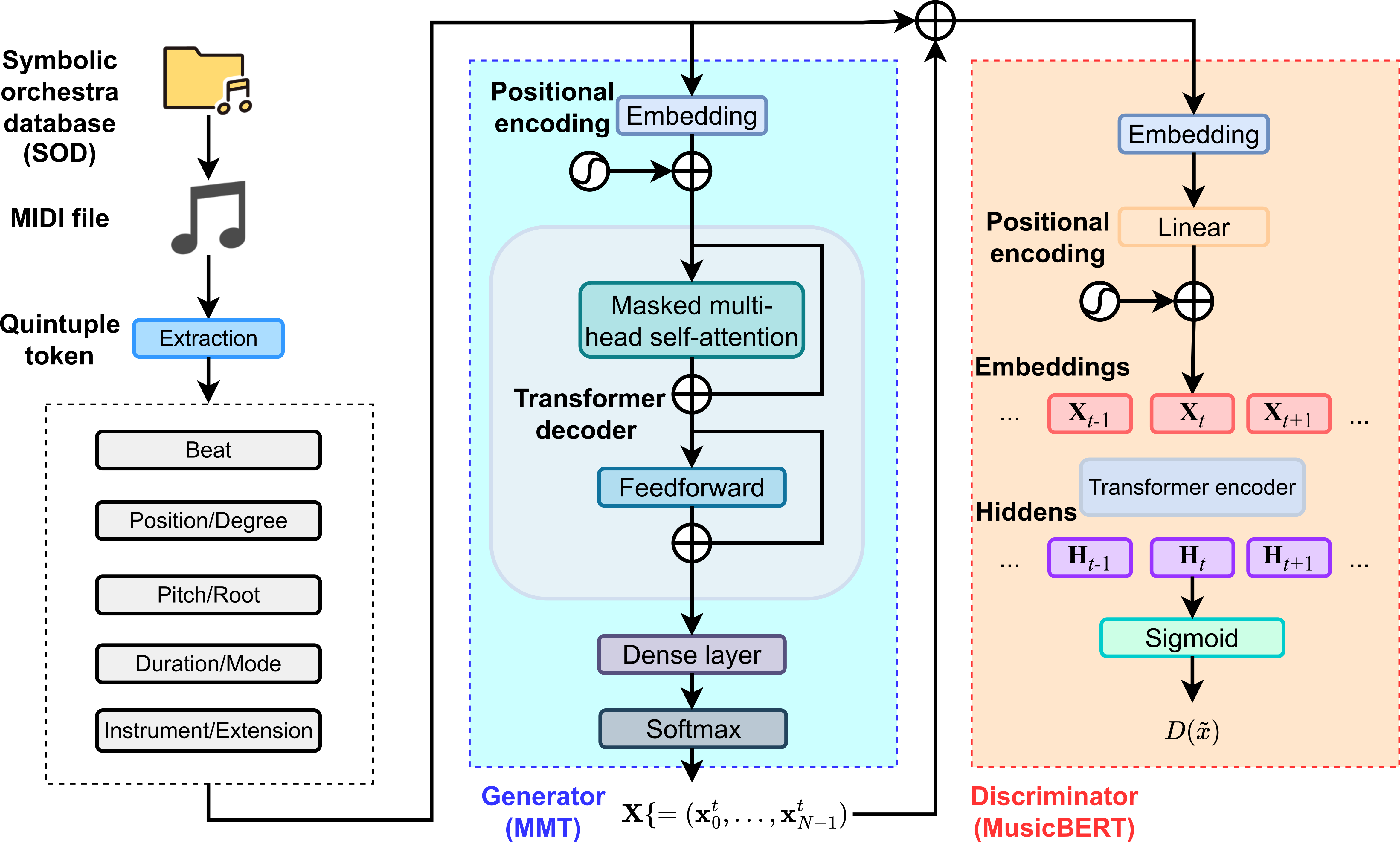}}
 \caption{The diagram of the MMT-BERT. The generator is built upon Transformer-XL architecture, and the discriminator is built upon MusicBERT. A MIDI file from the dataset is firstly encoded into a sequence of quintuple tokens before fed into the model. Embeddings of the 6 elements are concatenated by a linear layer and converted into a single vector. Then they are fed into the encoder and decoder layers with the addition of position embeddings.}
 \label{fig:flowchart}
 \vspace{-0.25cm}
\end{figure*}

\subsection{MMT-BERT Architecture}\label{subsec:Model-method}
The fundamental structure of our MMT-BERT architecture is based on a GAN architecture, employing MMT as the generator and MusicBERT as the discriminator.
The overview diagram of MMT-BERT is illustrated in~\figref{fig:flowchart}.
The primary concept of GAN is minimizing the loss to enhance the generator's ability to deceive the discriminator by producing fake music indistinguishable from real music, while simultaneously maximizing the discriminator's accuracy in distinguishing between real and fake music.
Details of the generator and discriminator will be discussed later.

\subsubsection{Generator}\label{subsubsec:G}
As the generator, we employ MMT~\cite{dong2023multitrack}, a Transformer-XL-based model that consists solely of decoders.
In MMT, elements in the quintuple token $\mathbf{x}_{i}$ are individually embedded first, and then concatenated, followed by the addition of positional embeddings. 
Subsequently, this combined input is passed through transformer decoder blocks, which are composed of a masked multi-head self-attention layer and a feedforward layer. 
The output from the decoder blocks is then processed by a dense layer and a softmax layer, resulting in the generation of new music samples.
MMT proves advantageous due to its capability to handle multi-dimensional input and output spaces, aligning perfectly with the requirements of symbolic music representation. 
Significantly, MMT can retain hidden states from previous segments, thereby eliminating the need for recalculating from scratch with each new segment. 
These retained states function as a memory aid for the current segment, establishing a recurrent connection between segments. 

The application of MMT as the generator allows for instrument-controllable multitrack music generation with extended duration and higher training speed.
Such a key feature facilitates the modeling of extensive long-range dependencies.

\subsubsection{Discriminator}\label{subsubsec:D}
As the discriminator, we adopted MusicBERT~\cite{zeng2021musicbert}, a large-scale Transformer model developed for symbolic music understanding.
MusicBERT consists of a Transformer encoder and utilizes a masked language modeling approach where certain tokens in the input music sequence are masked and then predicted by the model output.
The original proposed encoding method, called OctupleMIDI process transforms a symbolic music piece into a sequence of octuple tokens, each containing eight basic elements related to a musical note.
In order to make MusicBERT act as a discriminator adapted to the proposed representation mentioned in Sec.~\ref{subsec:SMR-method}, we refine the input and output format of MusicBERT. 
Quintuple tokens are converted into a single vector through the concatenation of embeddings and a linear layer. 
The resulting vector is combined with position embeddings and provided as input to the Transformer encoder.
To predict each of the five tokens within the quintuple, separate softmax layers are added to map the hidden states of the Transformer encoder to the vocabulary sizes of the different element types.
MusicBERT's proficiency in comprehending symbolic music as the discriminator integrates with MMT's generation process, thereby aiding in the stability of the training process and faster convergence.

\subsubsection{Relativistic Standard Loss}\label{subsubsec:rsloss}
Inspired by RS-GAN~\cite{li2023algorithm}, one of the state-of-the-art methods in GANs, we adopt the relativistic standard loss as our objective function. 
Applying relativistic standard loss prevents the network from becoming overconfident, leading to slower and more careful decisions, allowing the generator more room to adjust its weights and improve the training process~\cite{jolicoeur2018relativistic}.
The probability that the given fake data is more realistic than a randomly sampled real data is defined as follows:
\begin{equation}
    D(\Tilde{x}) = \mathrm{sigmoid}(C(f)-C(r)),  \label{rev}
\end{equation}
where $C(\cdot)$ denotes a non-transformed layer, and $\Tilde{x}$ denotes real/fake data pairs $\Tilde{x}=(r,f)$.
Hence, the loss function of the generator $G$ and the discriminator $D$ are defined as follows:
\begin{equation}\label{LG}
\begin{split}
        L_{G} = &\mathbb{E}_{(\mathbf{r},\mathbf{f}){\sim}p_{(r,f)}}[\log(\mathrm{sigmoid}(C(\mathbf{r})-C(\mathbf{f}))]\\
        &-\sum_{i}{r_i{\log{f_{i}}}},
\end{split}
\end{equation}
\begin{equation}\label{LD}
\begin{split}
    L_{D} = &\mathbb{E}_{(\mathbf{r},\mathbf{f}){\sim}p_{(r,f)}}[\log(\mathrm{sigmoid}(C(\mathbf{f})-C(\mathbf{r}))],
\end{split}
\end{equation}
where ${\it r_{i}}$ and ${\it f_{i}}$ denote ground truth logits and generated music logits, respectively. 
It is noted that we add cross entropy to the loss function of generator in order to accelerate the convergence process of the loss function.
By training both the generator and discriminator with the relativistic standard loss to emulate human-like musical compositions, our MMT-BERT model can generate high quality music pieces that incorporate sophisticated chord information.

\section{Experiment}\label{sec:Experiment}
\subsection{Experiment Setup}\label{setup}
In the experiment, we utilize the Symbolic Orchestral Database (SOD)~\cite{crestel2018database}, which comprises 5,864 music pieces encoded as MIDI files along with associated  metadata. 
The dataset is partitioned into training, testing, and validation sets, receiving 80\%, 10\%, and 10\% of the data, respectively.
We set a temporal resolution of 12 time steps per quarter note for detailed timing accuracy.
The Transformer-XL generator is composed of six decoder layers with 512 dimensions and eight self-attention heads, and the MusicBERT discriminator consists of one encoder layer with two self-attention heads.
The maximum length for symbolic music sequences is set at 1024, with a maximum of 256 beats. 
To optimize the models, we employ the Adagrad optimizer to mitigate issues of gradient explosion and vanishing~\cite{duchi2011adaptive}. 
Additionally, to enhance the robustness of the data, we augment it by randomly transposing all pitches by $s\sim{U}(-5,6)(s\in\mathbb{Z})$ semitones and assign a starting beat.
Here, ${U}$ denotes a uniform distribution.

As comparative methods, we employ three state-of-the-art music generation models: MMM~\cite{ens2020mmm}, FIGARO~\cite{von2022figaro}, and MMT~\cite{dong2023multitrack}. 
We validate the performance of our MMT-BERT model by conducting quantitative evaluations using existing metrics and subjective experiments to assess the human-like qualities of the generated music pieces.

\begin{table*}[t]
  \begin{center}
    \begin{tabular}{l c c c c c} \hline
       & PCES (\%) & SCS (\%) & GCS (\%) & AL (sec) \cr \hline 
       MMM\cite{ens2020mmm} & 92.93$\pm$1.22 & 98.64$\pm$0.92 & 98.28$\pm$0.29 & 38.69 \cr
       FIGARO\cite{von2022figaro} & 94.33$\pm$0.31 & 98.70$\pm$0.22 & \underline{98.84$\pm$0.67} & 28.69 \cr
      MMT\cite{dong2023multitrack} & 95.19$\pm$0.45 & 98.94$\pm$0.77 & 98.44$\pm$0.55 & \bf{100.42} \cr \hline \hline 
      MMT-BERT w/o Chord event& 95.57$\pm$1.32 & 98.81$\pm$0.23 & 99.56$\pm$0.32 & \underline{100.25} \\ 
      MMT-BERT w/o MusicBERT& \underline{96.22$\pm$0.44} & \underline{99.14$\pm$0.29}  &  98.61$\pm$0.44 & 97.43\\ 
            MMT-BERT (ours)& \bf{99.73$\pm$0.21} & \bf{99.64$\pm$0.31} & \bf{99.66$\pm$0.25} & 99.87 \\ \hline
    \end{tabular}
  \end{center}
  \vspace{-0.2cm}
  \caption{Quantitative evaluation results. The {\bf boldface} denotes the highest value, and the \underline{underlined} denotes the second highest value, respectively.}
  \label{objective}
\end{table*}

\begin{table*}[t]
  \begin{center}
    \begin{tabular}{l c c c c c} \hline
       & R & H & C & S & O  \cr \hline
             MMM\cite{ens2020mmm}& 3.83$\pm$0.92 & 3.78$\pm$0.87 & 3.78$\pm$0.73 & 3.67$\pm$0.84 & 3.83$\pm$0.79\cr
      FIGARO\cite{von2022figaro}& 3.78$\pm$1.11 & 3.78$\pm$1.11 & 3.89$\pm$1.02 & 3.89$\pm$1.13 & 3.83$\pm$0.92\cr \hline  \hline
       MMT\cite{dong2023multitrack} & 3.22$\pm$0.70 & 3.17$\pm$0.98 & \bf{3.44$\pm$1.03} & 3.33$\pm$1.09 & 3.22$\pm$0.78 \cr
      MMT-BERT (ours)& \bf{3.55$\pm$0.94} & \bf{3.55$\pm$0.92} & 3.33$\pm$0.98 & \bf{3.39$\pm$0.90} & \bf{3.44$\pm$0.80}\cr \hline

    \end{tabular}
    \end{center}
    \vspace{-0.2cm}
  \caption{Subjective evaluation results. Each metric is rated on a five-point scale, with the average score being calculated.}
  \vspace{-0.2cm}
  \label{subjective}
\end{table*}

\subsection{Quantitative Evaluation}\label{objecteval}
Following~\cite{dong2023multitrack}, we evaluate the generated music pieces using four metrics: pitch class entropy similarity (PCES), scale consistency similarity (SCS), groove consistency similarity (GCS), and average length (AL). 
We consider higher values of PCES, SCS, and GCS as indicators of superior quality, while a higher AL denotes a greater capability to produce long-duration music pieces.

In preparation for calculating PCES, the pitch class entropy (PCE) is defined as follows:
\begin{equation}
    \mathrm{PCE} = -\sum_{i=0}^{11}h_{i}\log_{2}(h_{i}),
\end{equation}
where $h_{i}$ denotes the number of occurrences of each note name in the 12-dimensional pitch class histogram.
As the PCE values increase, the tonality of the generated music pieces exhibits greater instability. 
However, it is important to recognize that more stable tonality does not necessarily imply higher quality.
Subsequently, we calculate PCES between generated music samples and human compositions as follows:
\begin{equation}\label{PCES}
    \mathrm{PCES} = 1-\frac{|\mathrm{PCE}_{\mathrm{gen}}-\mathrm{PCE}_\mathrm{tr}|}{{\mathrm{PCE}_{\mathrm{tr}}}},
\end{equation}
where $\mathrm{PCE}_{\mathrm{gen}}$ and $\mathrm{PCE}_{\mathrm{tr}}$ denotes the PCE value of generated music samples and human compositions, respectively. 
Moreover, noticing that PCE is intrinsically linked to the volume of data, we truncate the generated musical pieces to the preceding $k$ seconds and calculate their PCES.

The scale consistency (SC) is derived by calculating the proportion of tones that conform to a conventional scale and presenting the value for the most closely aligned scale~\cite{mogren2016c}. 
SC serves as an indicator of the model's proficiency in generating musical segments that demonstrate cognizance of chords and scales within the current bar. 
The SCS between generated music samples and human compositions is defined as follows:
\begin{equation}\label{SCS}
    \mathrm{SCS} = 1-\frac{|\mathrm{SC}_{\mathrm{gen}}-\mathrm{SC}_\mathrm{tr}|}{{\mathrm{SC}_{\mathrm{tr}}}},
\end{equation}
where $\mathrm{SC}_{\mathrm{gen}}$ and $\mathrm{SC}_{\mathrm{tr}}$ denote the SC values of generated music samples and human compositions, respectively.

To calculate GCS, we first define a groove pattern $\mathbf{g}$ as a 64-dimensional binary vector.
The groove consistency (GC) between two grooving patterns $(\mathbf{g}^{a},\mathbf{g}^{b})$ is defined as follows:
\begin{equation}
    \mathrm{GC} = 1 - \frac{1}{Q}\sum_{i=0}^{Q-1}\mathrm{XOR}(g_{i}^{a},g_{i}^{b}),
\end{equation}
where XOR$(\cdot,\cdot)$ denotes the exclusive OR operation, and $g_{i}$  denotes a position in a bar at which there is at least a note onset.
$Q$ is the dimensionality of $\mathbf{g}^{a}$ and $\mathbf{g}^{b}$.
GC is a measure of music's rhythmicity. The value of GC stands for the steadiness in rhythm of the generated music pieces.
The GCS between generated music samples and human compositions is defined as follows:
\begin{equation}\label{GCS}
    \mathrm{GCS} = 1-\frac{|\mathrm{GC}_{\mathrm{gen}}-\mathrm{GC}_\mathrm{tr}|}{{\mathrm{GC}_{\mathrm{tr}}}},
\end{equation}
where $\mathrm{GC}_{\mathrm{gen}}$ and $\mathrm{GC}_{\mathrm{tr}}$ denote the GC values of generated music samples and human compositions, respectively.

AL denotes the mean duration of the generated music pieces, which collectively illustrates the model's ability to generate musical sequences with significant length.

The results of the quantitative evaluation are shown in~\tabref{objective}.
To facilitate a fair comparison by standardizing the lengths of music pieces, PCES is assessed over a 15-second span due to the limitations of MMM and FIGARO in producing extended compositions.
Experimental results show that MMT-BERT achieves higher performance in PCES, SCS, and GCS compared to the other methods, demonstrating its effectiveness in generating high quality music pieces.
This achievement is attributed to its chord awareness and the symbolic music understanding facilitated by MusicBERT.
MMT-BERT's AL is marginally less than that of MMT, and this results from integrating chord events that are not converted to audio during the decoding phase.
However, MMT-BERT's AL significantly surpasses that of MMM and FIGARO, confirming its capability to generate longer compositions. 
Additionally, the AL of all the music pieces in the SOD we used, which also serve as the ground truth, is 99.88 seconds.
Evaluation results show that MMT-BERT can produce music of higher quality than MMT, and of longer duration than MMM and FIGARO.

\subsection{Impacts of Chord Event and Discriminator}\label{subjectiveeval}
MMT-BERT aims to generate more harmonious, more human-like music pieces through the addition of chord events and adversarial generative learning by employing MusicBERT as its discriminator. 
To evaluate aspects related to richness and humanness, we have conducted subjective experiment and ablation study.

In the subjective experiment, we asked 18 music amateurs as the following five questions and requested that they rated each on a five-point scale.
\begin{itemize}
\item \textbf{Richness (R)}: Does the music piece have diversity and interestingness?
\item \textbf{Humanness (H)}: Does the music piece sound like it was composed by an expressive human musician?
\item \textbf{Correctness (C)}: Does the music piece contain perceived mistakes in composition or performance?
\item \textbf{Structureness (S)}: Does the music piece exhibit structural patterns such as repeating themes or the development of musical ideas?
\item \textbf{Overall (O)}: What is the general score of the music piece?
\end{itemize}
As mentioned in Sec.~\ref{objecteval}, FIGARO and MMM employ a music representation that considers percussive sounds and typically generates much shorter pieces.
Therefore, the nature of the music pieces generated by these models, FIGARO and MMM, differs significantly from that of MMT-BERT and MMT due to their use of percussive sounds and shorter compositions. 
To fairly evaluate the human-like quality of the generated music pieces, we compared MMT-BERT with MMT, a state-of-the-art approach whose generated compositions have lengths and musical styles that are relatively similar to those of MMT-BERT.
Additionally, to ensure clarity in subjective evaluation, we included the results for MMM and FIGARO.
The results of the subjective evaluation are shown in~\tabref{subjective}. 
\tabref{subjective} indicates that MMT-BERT scores are particularly high in both richness and humanness compared to MMT.
This suggests that the application of chord events and MusicBERT contribute to the generation of music pieces that more closely resemble human compositions.
On the other hand, regarding correctness, our method did not specifically aim to enhance this metric, which may cause the gap in this value.
For the same reason, our method exceeds MMT by a small margin in structureness mainly because of uncertainty.
Although there is no clear advantage between MMT and MMT-BERT in correctness and structureness, our method still outperforms MMT in richness and humanness. 
The overall score also proves the superiority of MMT-BERT, which indicates that chord events and MusicBERT enhance the ability to create music similar to that produced by humans.

The results of the ablation study are shown in~\tabref{objective} along with the quantitative evaluation results. It is evident that the addition of chord events
improves PCES and SCS. MusicBERT also contributes to the enhancement of PCES and GCS.

\section{Conclusion}\label{conslusion}
In this paper, we have proposed the chord-aware symbolic music generation approach, named MMT-BERT. 
By extracting chord information from raw audio files, we have devised a chord-aware symbolic music representation. 
We also developed a novel RS-GAN architecture based on MMT and MusicBERT. 
Both experimental evaluations validate the efficacy of our method in producing music pieces of superior quality, enhanced human likeness, and considerable length.
In future works, we plan to explore methods that refine musical structure and incorporate information from various musical modalities.

\section{Acknowledgments}
This work was partly supported by JSPS KAKENHI Grant Numbers JP21H03456, JP23K11141 and JP23KJ0044.

\bibliography{ISMIR2024}

\end{document}